\documentclass[twocolumn]{aa}
\usepackage{natbib,twoopt}
\usepackage[breaklinks=true]{hyperref} 
\usepackage[varg]{txfonts}
\bibpunct{(}{)}{;}{a}{}{,} 
\usepackage[utf8]{inputenc}
\usepackage[english]{babel}
\usepackage{amsmath}
\usepackage{amsfonts}
\usepackage{amssymb}
\usepackage{graphicx}
\usepackage{lmodern}
\usepackage{tikz}
\usepackage{hyperref}
\usepackage{color}

\begin{document}

\title{Timing offset calibration of CZTI instrument aboard ASTROSAT}
\author{Avishek Basu \inst{1} \and Bhal Chandra Joshi\inst{1} \and Dipankar Bhattacharya\inst{2} \and A R Rao\inst{3} \and A. Naidu\inst{1,}\inst{4} \and M. A. Krishnakumar\inst{1,}\inst{6} \and Prakash Arumugsamy \inst{1} \and Santosh Vadawale\inst{5} \and P.K. Manoharan\inst{1,}\inst{6} \and G.C. Dewangan\inst{2} \and Ajay Vibhute\inst{2} \and N.P.S. Mithun\inst{5} \and Vidushi Sharma\inst{2}}
\institute{National Centre for Radio Astrophysics-Tata Institute of Fundamental Research, Pune, India
\and The Inter-University Centre for Astronomy and Astrophysics, Pune, India
\and Tata Institute of Fundamental Research, Mumbai, India
\and McGill Space Institute, McGill University, Montreal, Canada
\and Physical Research Laboratory, Ahmedabad, Gujarat, India
\and Radio Astronomy Centre, National Centre for Radio Astrophysics-Tata Institute of Fundamental Research, Udagamandalam, India
}
\date{Received $27/02/2018$ /
Accepted $29/05/2018$}

\abstract{ \textit{Aim}: Both the radio and the high-energy emission mechanism in pulsars is not yet properly understood. A multiwavelength study is likely to help better understand of such processes.  ASTROSAT, the first Indian space-based observatory,  has five instruments aboard that cover the electromagnetic spectrum from infra-red (1300 $\AA$) to hard X-ray (380 keV). The instrument relevant
to our study is   the Cadmium Zinc Telluride Imager (CZTI). CZTI is a hard X-ray telescope functional over an energy range of 20$-$380 keV. We aim to estimate the timing offset introduced in the data acquisition pipeline of the instrument, which will help in time alignment of high energy time-series with those from two other ground based observatories, viz. the Giant Meterwave Radio Telescope (GMRT) and the Ooty Radio Telescope (ORT). \\
\\ \textit{Method}: PSR B0531+21 is a  well studied bright pulsar with closely aligned radio and hard X-ray pulse profiles. We used simultaneous observations of this pulsar with the ASTROSAT, the ORT, and the GMRT. As the pulsar resides in a very turbulent environment and shows significant timing noise, it was specially observed using the ORT with almost daily cadence to obtain good timing solutions. We also supplemented the ORT data with  archival FERMI data for estimation of timing noise. We obtained the phase connected timing solution of the pulsar by estimating its dispersion measure variations and the timing noise. The timing offset of ASTROSAT instruments was estimated from fits to pulse arrival time data at the ASTROSAT and the radio observatories. \\
\\ \textit{Results}: We estimate the relative offset of ASTROSAT-CZTI with respect to GMRT to be -4716 $\pm$ 50 $\mu s$. The corresponding offset with the ORT was -29639 $\pm$ 50 $\mu s$ and Fermi-LAT was -5368 $\pm$ 56 $\mu s$ respectively.}

\maketitle

\section{Introduction}
Simultaneous multiwavelength timing and spectral observations are very important in studying emission mechanisms of 
astrophysical objects with high energy emission, such as  X-ray 
pulsars and X-ray binary systems \citep{YadiRomani95,MuslimovHarding03}, to model their phase resolved spectra \citep{HU17},
and to constrain their magnetic field structure. 
Many of objects, particularly pulsars with known radio pulsations, require 
high-precision alignments of their radio and high energy light curves. 
Apart from constraining the nature of pair producing gaps, high 
precision alignment of their light curves can shed light on the 
nature of giant pulses (GPs)\footnote{Intense 
nano second wide pulses, with typical intensities about 1000 times the mean 
pulse intensity, seen sporadically at radio frequencies in PSR B0531+21 
and some other pulsars} in some of these pulsars  \citep{Joshi2004,JohnstonRomani04,JhRomanireview04,lundgren1995giant,mikami2014,hankins2007radio}.

Unfortunately, the behavior of electromagnetic radiation makes it 
impossible to observe all bands using a single instrument. In radio 
telescopes, the data are recorded in the form of voltages as a function 
of time, whereas high energy detectors count the photons. The variability 
of both the radio and high energy emission necessitates having accurate  
synchronization of time, when the radiation at different frequency 
arrives at different observatories. This requires calibrating the delays 
in data acquisition and processing pipelines for each observatory 
through observations of sources with known time alignment. The 
calibration of fixed offsets for the instruments on board the 
first Indian multiwavelength space observatory, ASTROSAT, are presented 
in this paper for the first time.

ASTROSAT, launched in October 2015, has five instruments on board \citep{singh2014astrosat}. These are the  
Cadmium Zinc Telluride Imager  \citep[\textit{CZTI};][]    {bhalerao2016cadmium},  
Large Area X-ray Proportional Counter  
\citep[\textit{LAXPC};][]{yadav2016large}, 
Soft X-ray Telescope \cite[\textit{SXT};][]{singh2016orbit}, Ultra Violet 
Imaging Telescope  \citep[\textit{UVIT};][]{hutchings2014uvit}, and Scanning Sky Monitor (\textit{SSM}).  This is a unique observatory providing multiwavelength coverage from 1300 $\AA$ to 380 keV.

\begin{figure}
\label{radio_profile}
\includegraphics[scale=0.24]{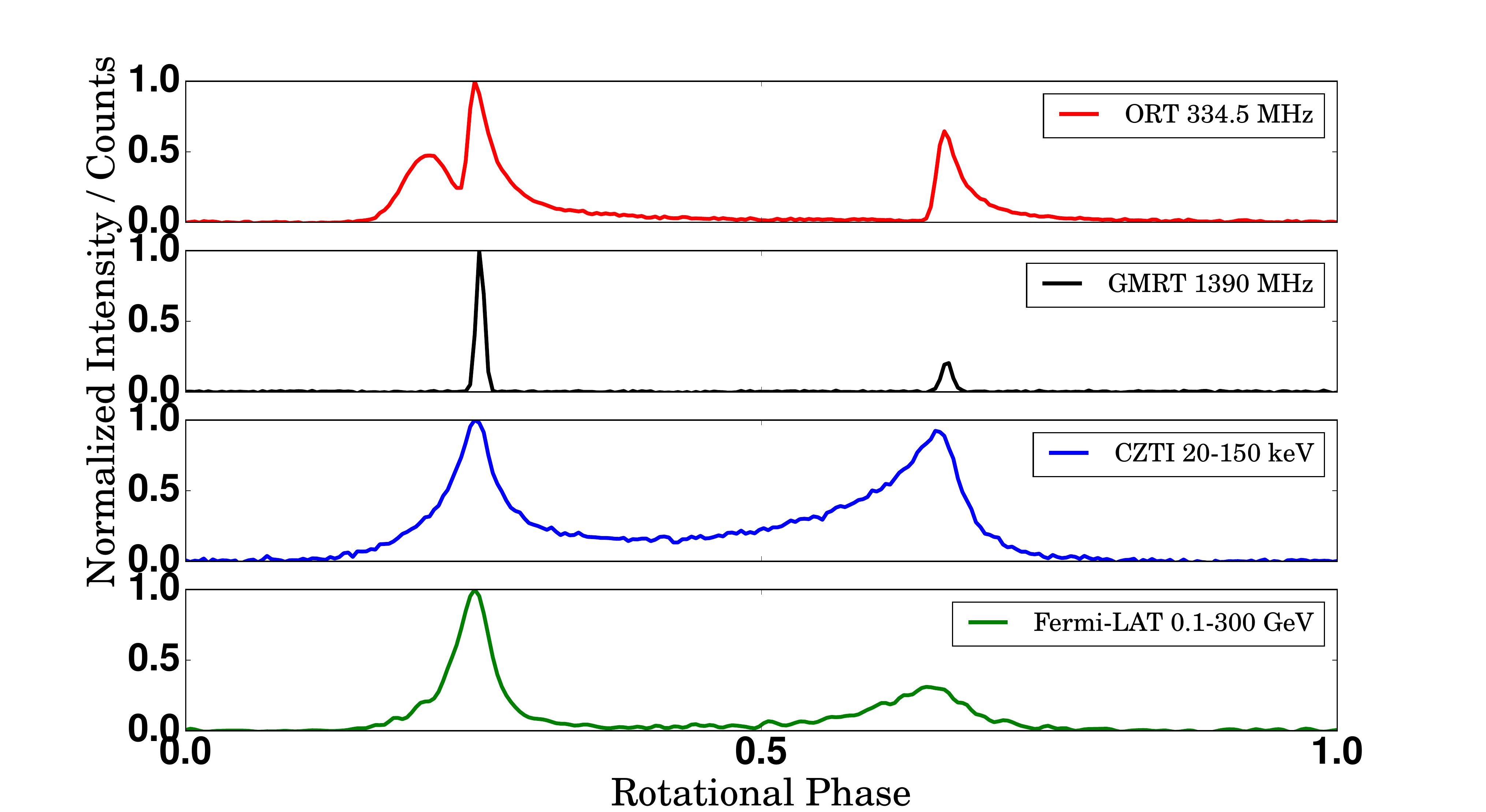}
\caption{Average pulse profile of PSR B0531+21 obtained using phase coherent average over all data with different instruments used in this study. The data were aligned using the offsets estimated in this study. The larger peak is called as main-pulse (MP), while the smaller peak is referred as inter-pulse (IP).
The panels show the average profiles obtained with FERMI archival data, the CZTI, the GMRT and the ORT from bottom panel to top respectively. The radio profile were obtained with the GMRT at 1390 MHz and with the ORT at 334.5 MHz.}
\end{figure}

We have used Indian ground-based facilities such as the Giant Meterwave Radio Telescope \citep[\textit{GMRT};][]{swarup1991giant} and the Ooty Radio Telescope \citep[\textit{ORT};][]{swarup1971large} simultaneously with the ASTROSAT to calibrate the fixed timing offsets of various instruments. For this, we needed to use a standard calibrator with well known properties in all bands. We used the brightest high energy pulsar in the sky, the Crab pulsar (PSR B0531+21), for our calibration. PSR B0531+21 emits pulsed radiation from radio to very high energies   \citep{abdo2010first}. The average light curve of the 
pulsar shows two peaks, with the larger peak at 1.4 GHz defined as 
the main pulse (\figurename{ \ref{radio_profile}}).

The pulsar's profile at low frequencies (334.5 MHz and 1.4 GHz) are aligned with the profile at optical and high energies. The main pulse at high energies leads the radio main pulse by 
241 $\pm$ 29 $\mu$s \citep[$>$ 30 Mev ;][]{kuiper2003absolute}, 344 $\pm$ 40 $\mu$s  \citep[2-30 keV ;][]{rots2004absolute} and (280 $\pm$ 40 $\mu$ s) \citep{kuiper2003absolute}. We used these reported intrinsic offsets to calibrate the instrumental offset of the instruments aboard the ASTROSAT.

This paper is organized as follows. The instruments aboard the ASTROSAT and the details of observations used for this study are described in Section \ref{obs}.  The ephemeris for PSR B0531+21 were obtained using high cadence radio observations at 334.5 MHz with the Ooty Radio Telescope (ORT). The analysis of these data and our calibration method is discussed in Section 
\ref{anal}. We conclude with results and discussion in Section \ref{result}. 

\section{Observations}
\label{obs}

The radio observations used the ORT and the GMRT, whereas the X-Ray observations were carried 
out using the CZTI instrument aboard ASTROSAT. We also used publicly available archival data from Fermi\footnote{https://fermi.gsfc.nasa.gov/ssc/data/access/}mission. These instruments and the observational setup along-with the details of observations are described in this section.

\subsection{The Ooty Radio Telescope (ORT) }
\label{ort}

The ORT is an offset parabolic 
cylindrical antenna of 530-m length in north-south direction and 
30-m width in east-west direction, sensitive to a single 
linear polarization,  with system temperature of 
150 K and the antenna gain of 3.3 K/Jy \citep{swarup1971large}. 
PSR B0531+21 was observed as part of a larger pulsar-monitoring program \citep{kjm+18} since 2014 March 
at 334.5 MHz with a bandwidth 
of 16 MHz. The pulsar was observed daily  for 15 minutes as part of this program 
and these observations were used to obtain monthly ephemeris of the 
pulsar as described below. The observations utilized the pulsar 
back-end at the ORT, called PONDER \citep{pondernaidu}, which employed 
real-time coherent dedispersion to obtain directly time-stamped average 
profiles of the pulsar using the monthly ephemeris generated by us. 
In PONDER, data acquisition is started at the rising edge of 
the minute pulse derived from global positioning system (GPS) and  
data are sampled in synchronization with observatory frequency standard, 
which was a Rubidium clock. The typical instrumental uncertainty 
on the time stamp was 200 ns. Observations from 2015 September 01 
(MJD 57226) to 2017 January 14 (MJD 57767) were used in this work.

\subsection{The Giant Meterwave Radio Telescope (GMRT)}
\label{gmrt}

The GMRT is an interferometer consisting of thirty 45-m fully 
steerable antennas \citep{swarup1971large}, 14 of which are arranged 
in a compact array within 1 km and the rest are distributed in three 
arms. We used the arm antennas nearest to the compact array along with the 14 compact 
array antennas to form a phased array at 1390 MHz, with an overall 
gain of 3.5 K/Jy. The two linear polarizations across 16 MHz bandwidth 
from each antenna were digitized at Nyquist rate. The resultant 
time series was transformed with a 512 point fast Fourier 
transform (FFT) to obtain 256 channel voltages in the frequency 
domain. These were then compensated in the Fourier domain  
for instrumental phase for each antenna, determined by observing a 
point source (3C147) before each observations. The phase compensated 
voltages from all antenna in the phased array were then added 
and this coherent sum was recorded as 256 channel complex voltages, 
with a time-stamp for each block of 256 channels derived from 
observatory Rubidium frequency standard disciplined using one pulse per minute 
output of GPS. The recorded voltages were processed offline 
as described in Section \ref{anal}. PSR B0531+21 was observed at the GMRT and the ORT simultaneously with ASTROSAT observations at four epochs. At other 13 epochs, the GMRT observations were not possible, so the ASTROSAT observations were carried out with the ORT only.
The details of observations used are given 
in Table \ref{obsdet}.

\begin{table*}[ht]
\label{obsdet}
\begin{tabular}{|l|l|l|l|}
\hline
{\bf Telescope}& {\bf BW or Energy range}& {\bf Start MJD}& { \bf Stop MJD}\\
\hline
ORT& 16 MHz& 57226& 57767\\
\hline
GMRT legacy& 16 MHz& 57316& 57772\\
\hline
ASTROSAT-CZTI& 20-150 keV& 57303& 57771\\
\hline
Fermi-LAT& 0.1-300 GeV& 57284&57800\\
\hline
\end{tabular}
\\
\caption{Brief summary of the observations. The participating telescopes and their payloads,  bandwidth or the energy range employed for observations, and the range of MJD for which the data have been used are listed. } \label{obsdet}
\end{table*}

\subsection{ASTROSAT-CZTI}
\label{czt}

The Cadmium Zinc Telluride Imager (CZTI) instrument \citep{bhalerao2016cadmium} aboard ASTROSAT is a two-dimensional coded mask Imager with solid state pixelated Cadmium Zinc Telluride detectors of 976 cm$^2$ total geometric area divided into four quadrants, each containing 4096 pixels.  The instrument operates in the energy range 20--150 keV for direct imaging, providing an angular resolution of $\sim 8$~arc-min within a field of view of $4.6^{\circ} \times 4.6^{\circ}$.  Events recorded by the CZTI are time stamped with a resolution of 20 $\mu$s as per the instrument clock. On ASTROSAT, the primary time standard is provided by a spacecraft positioning system (SPS) which generates a GPS-synchronized UTC reference. A synchronizing pulse is sent to all X-ray payloads once in every 16 UTC seconds. The
local clock values of all the instruments and of the SPS are recorded at each such pulse into a time correlation table (TCT).  Events recorded in the CZTI are assigned UTC time-stamps by interpolation in the TCT.  Accuracy of absolute time stamps thus assigned to CZTI events are estimated to be within $\sim 3 \mu$s standard deviations  \citep{bhattacharya2017}.  Unlike most other space observatories, the event time stamps in ASTROSAT are provided in the UTC system instead of TT.  In order to derive the Barycentric arrival time of each event, these UTC time-stamps are processed, along with information regarding the orbital motion of ASTROSAT, through a modified version of the well-known AXBARY task of NASA HEASOFT package.  The modification, made available under the name "as1bary",  takes into account the additional bookkeeping required for leap seconds while processing UTC time stamps.

\section{Analysis}
\label{anal}

\subsection{Analysis of radio data}
\label{analrad}

As mentioned in Section \ref{ort}, the data obtained at the ORT 
were already available as coherently dedispersed time-stamped profiles, 
which were used in the timing analysis described later. The GMRT 
spectral voltage data were coherently dedispersed offline using a 
pipeline developed by us. This pipeline first converts the spectral 
voltages to a voltage time series by taking an inverse FFT. The 
time-series is then convolved with a unity gain phase delay 
filter, representing the effect of  inter-stellar medium 
as described in \cite{pondernaidu}. Both the coherently 
dedispersed time series as well as an average profile, folded 
using the monthly ephemeris generated with the ORT data translated to the start time of the observations, were 
recorded for further analysis after integration to a resolution of 
1 $\mu$s.  The dispersion measure\footnote{Dispersion measure 
is the integrated column density of electrons in the line 
of sight of the pulsar, expressed in units of pc\,cm$^{-3}$} (DM) 
used in the real-time coherent dedispersion carried out in 
PONDER back-end at the ORT used the DM value provided in the 
Jodrell Bank monthly 
ephemeris\footnote{http://www.jb.man.ac.uk/pulsar/crab.html} \citep{lyne199323} 
nearest to the epoch of 
observations. As the GMRT data were coherently dedispersed 
offline, subsequent to observations, DM derived from our 
timing analysis was used in this case to obtain the folded average 
profiles. The profiles, obtained with the ORT and the GMRT, were converted 
to PSRFITS\footnote{PSRFITS is an open data storage format, 
which is based on the flexible image transport system 
(FITS) \citep{hsm04}} format. 

In addition to pulse profiles, the off-line analysis of the GMRT data 
also yielded data dedispersed to 64 sub-bands within the 16 MHz band-pass. 
These were folded to 32 sub-integration for each of the 600 s observations at the 
GMRT and converted to PSRFITS.

First, a noise-free template was created from the observed 
average profiles for a given telescope as described below. The pulse 
at 334.5 MHz is broadened due to multipath propagation in the inter-stellar 
medium (\figurename{ \ref{radio_profile}}). Furthermore, 
the pulse suffers a variable scatter-broadening  at this frequency 
due to varying inhomogeneities in the Crab nebula. This 
can introduce a systematic error into the estimation of time-of-arrivals (TOAs) depending 
on the extent of scatter-broadening in the profiles used for forming 
the template. Hence, we chose a high signal-to-noise ratio (S/N) 
profile with minimum scatter broadening for 
creating a template at this frequency. As the scatter-broadening 
is negligible at 1390 MHz (\figurename{ \ref{radio_profile}}), 
the template at this frequency was obtained from a 
profile generated by aligning and averaging best average profiles from several epochs of 
observations. These profiles were then modeled as a sum of Gaussians, 
using tools in PSRCHIVE package \citep{hsm04}, to obtain a noise free templates. 
Separate templates were obtained for the ORT and the GMRT and these 
were aligned with the MP positioned at pulse phase 0.24.

The average profile for each epoch at a given frequency were 
cross correlated with the noise free template for that frequency
using a  Fourier domain method \citep{taylor1992pulsar} to obtain the shift at 
each epoch. The time-stamp at each epoch was adjusted by 
this shift to obtain TOA. 
These TOAs were used to refine the pulsar rotation parameter using pulsar timing package 
TEMPO2 \citep{hobbs2006tempo2} \footnote{http://www.atnf.csiro.au/research/pulsar/tempo2/}. 
In brief, this technique (called pulsar timing) compares 
the observed TOAs with those predicted by an assumed  rotation 
model of the pulsar, keeping track of every rotation and minimizing the timing residuals through a least square fit 
PSR B0531+21 shows rotational irregularities in the form 
of timing noise \citep{scotttimingnoise03} and 
glitches \citep{crabglitchwang12}. 
These irregularities can significantly affect the residuals 
leading to phase ambiguities. Thus, closely spaced observations 
of pulsar are required to keep track of pulse phase and maintain 
phase connection. Our experiment used the ORT for high cadence 
observations of the pulsar to achieve the required phase 
connection.

\begin{table}[h]
\begin{tabular}{|l|l|}
\hline
Pulsar parameter     & Value         \\ \hline
RAJ  (hh:mm:ss)      &  05:34:31.973 \\
DECJ (dd:mm:ss)      &  +22:00:52.06 \\
F0   (Hz)            &   29.6607409(4) E$-$7 \\
F1   (Hz s$^{-1}$)    &   $-$3.6937842(9) E$-$10 \\
F2   (Hz s$^{-2}$)    &   1.1905(3) E$-$20  \\
PEPOCH (MJD)         &  57311.000000136 \\
POSEPOCH (MJD)       &  40675 \\
DMEPOCH  (MJD)       &  57311.000000136 \\
DM    (pc\,cm$^{-3}$)  &  56.7957 \\
PMRA  (mas/year)     &  $-$14.7 \\
PMDEC (mas/year)     &  2 \\
WAVE\_OM  (year$^{-1}$)  &   0.0054325986245627 \\
WAVEEPOCH (MJD)  &  57311.000000136 \\
DMMODEL (pc\,cm$^{-3}$)   &  56.7957 \\ \hline
\end{tabular}
\\
\caption{ Reference timing solution for PSR B0531+21 
after accounting for the timing noise and DM variations using the 
multiband observations presented in this paper.}

\label{timsol}
\end{table}

The TOAs from the ORT data were divided in 30 day intervals and 
local fits to the spin frequency (F0) and its derivatives (F1 and F2) 
were performed at an epoch in the center of each 30-day 
interval. These 30-day ephemerides were then used for folding 
the high energy data as well as the 1390 MHz GMRT data. The details 
of full phase connected timing analysis are described in Section \ref{analtim}.

\begin{figure*}[ht]
\includegraphics[scale=.4]{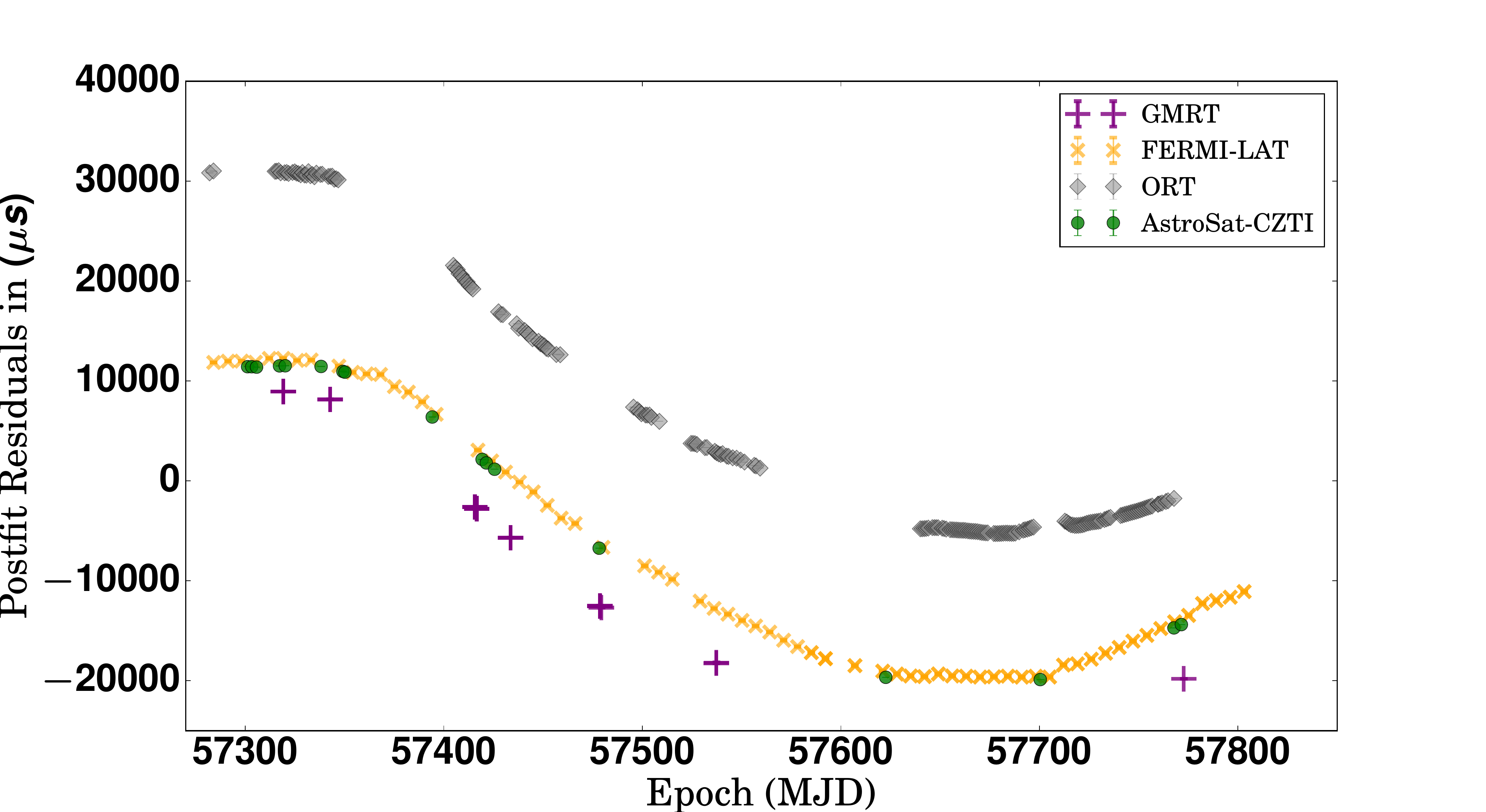}
\caption {Phase connected TOAs from the Fermi-LAT (yellow cross markers), the ASTROSAT CZTI (green circles), the GMRT (purple plus markers) and the ORT (gray diamonds) observations. The phase connection was obtained with the high cadence TOAs derived from the ORT observations and then applied to TOAs from other telescopes. The systematic pattern in the timing residuals is due to timing noise. The TOAs for different telescopes are offset with each other due to relative delays in the data acquisition at each telescope.}
\label{ortphsplt}
\end{figure*}

\subsection{Analysis of high energy data}
\label{analhigh}

\subsubsection{Analysis of Fermi-LAT data}
\label{analhighfermi}

We used the available archival data from 
Fermi-LAT\footnote{https://fermi.gsfc.nasa.gov/cgi-bin/ssc/LAT/LATDataQuery.cgi} 
and extracted all events in a 3-deg radius around the 
position of PSR B0531+21 in the energy range of 0.1 to 300 GeV. 
These were then split into separate event files, each spanning 
seven days using Fermi science 
tools\footnote{https://fermi.gsfc.nasa.gov/ssc/data/analysis
/scitools/overview.html}. The event times were referenced to the solar system barycenter 
(SSB) and the events were folded using the Fermi 
plugin \citep{rkp+11} of TEMPO2 with the ephemeris obtained in 
Section \ref{analrad}. A template for the averaged 
light curve in gamma-ray energies was constructed in a 
manner similar to the radio data and was aligned with the 
1390 MHz and the 334.5 MHz templates. The TOAs for each seven-day integrations 
were then derived by cross-correlating with this template and used 
in the subsequent timing analysis.

\subsubsection{Analysis of ASTROSAT data}
\label{analhighasat}

As mentioned before, instruments on board ASTROSAT provide 
individual photons with time-stamps derived from a satellite 
positioning system (SPS). The time tags of the photon were  
converted to solar system barycentre using the position 
of satellite in a code called {\it as1bary}.  
The barycentered events were then binned across 256 pulse phase bins 
using the ephemeris obtained in Section \ref{analrad}.  The binned profile were then written as PSRFITS files.

\begin{figure}[h]
\includegraphics[scale=0.22]{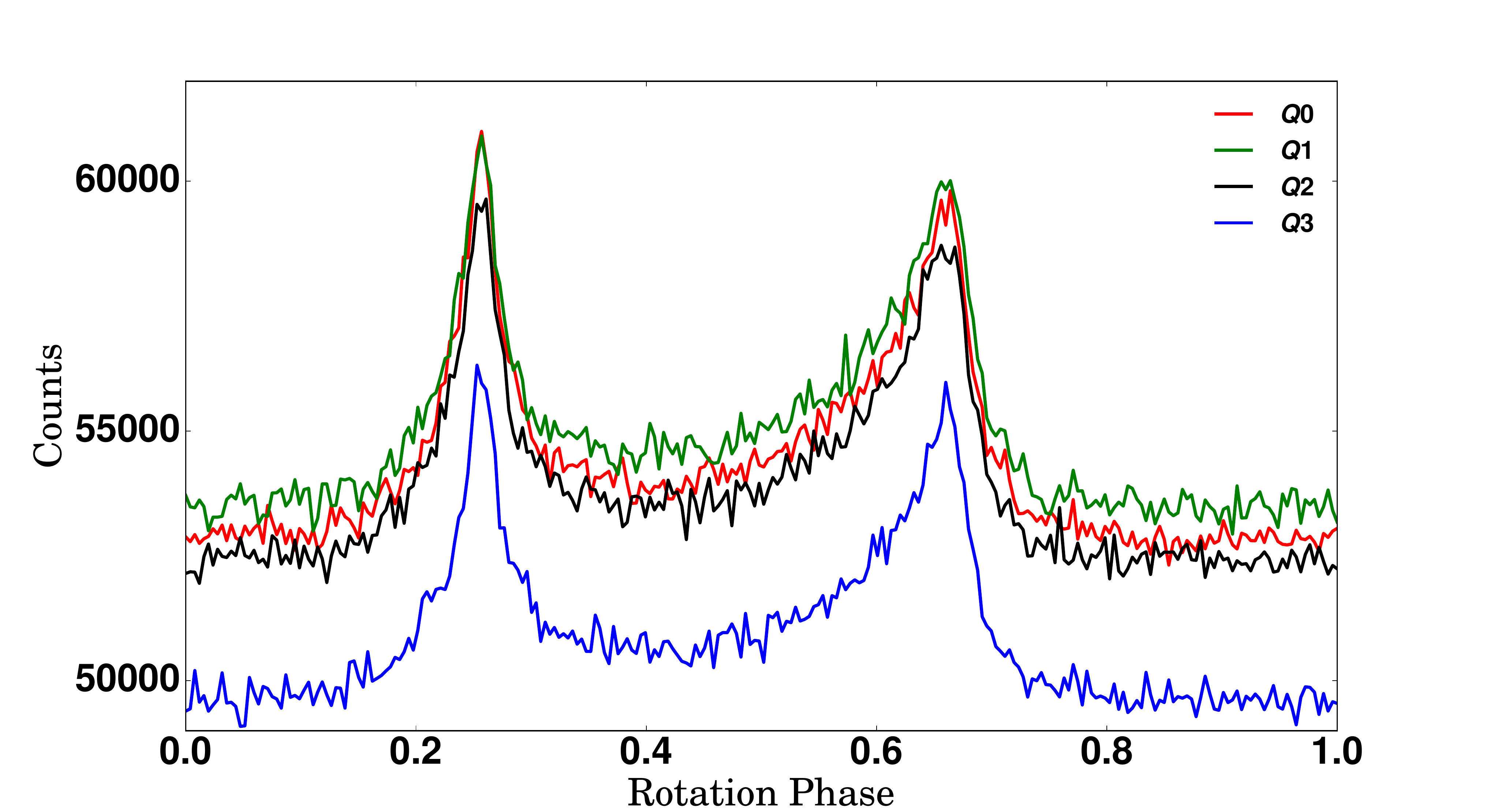}
\caption{Phase aligned profiles of PSR B0531+21 using four different detectors of CZTI arranged in a four quadrant fashion (Q0, Q1, Q2 and Q3)}
\label{cztquad}
\end{figure}

CZTI instrument has four quadrant detectors. Hence for a good S/N, 
we needed to combine data from all the quadrants. 
We checked the alignment of the individual detector 
data by  folding the photons from each quadrant separately as well 
as after combining the data from all quadrants. The profiles, 
so obtained, are shown in \figurename{  \ref{cztquad}}, where the phases were 
appropriately aligned using TEMPO2.  All the analysis in 
this paper uses the data combined from all four quadrants.

Separate profile templates were  constructed for CZTI data  
in a manner similar to the radio template. These were aligned with the 
the Fermi-LAT, the GMRT and the ORT templates. Finally, the CZTI  
template was cross-correlated with the observed profiles for CZTI to obtain TOAs in a manner similar to Fermi data. These were 
subsequently used in the timing analysis described in the next section.

Timing offsets evaluated from observed differences in the Crab pulsar phase may suffer from ambiguities amounting to integral multiples of the pulse period.  To test whether the offset between AstroSat-CZTI and Fermi could be as large or larger than 33 milliseconds, we compared the detection times of gamma ray bursts by these two missions.  In particular, the bright, short burst GRB170127C \citep{bissaldi,vidushi} provided the best S/N for this test.  We binned the UTC light curves from Fermi-GBM and AstroSat-CZTI at 10~ms resolution. The cross-correlation function of these two light curves showed a sharp peak at a delay of 0.0~ms with a formal error of 2.3~ms ($1\sigma$).  The relative distance between the two spacecrafts, projected in the direction of the GRB, was 877~km at the time of this detection, corresponding to a travel time difference of 2.9~ms. We therefore conclude that the difference in the absolute time stamps of Fermi and AstroSat-CZTI is much less than 33~ms and hence no integral-period ambiguity is expected in the relative phase comparison of the Crab pulsar between these two missions.

\subsection{Timing analysis }
\label{analtim}

All the radio and high energy TOAs, analyzed using 
the high cadence timing solution obtained with the ORT, are 
shown in \figurename{ \ref{ortphsplt}}. The timing noise is 
clearly visible in this plot and so are the relative offsets 
between the telescopes. The assumed parameters of the timing model are given 
in Table \ref{timsol} along-with the reference epochs.

The timing analysis was done using the pulsar-timing 
package TEMPO2. First, a reference timing solution was 
obtained by local fits to ORT high cadence TOAs between 
MJD 57282$-$57324 with a model involving the known 
astrometric and rotational parameters and DM for the pulsar. 
The fitted ephemeris were then used to phase connect 
the TOAs of all the telescopes as shown in \figurename{\ref{ortphsplt}}. This reference ephemeris was the 
starting point for the subsequent analysis described below.

The main objective of this work was to estimate the offset  
in data acquisition pipeline of ASTROSAT. This was done 
by comparison of the phase of the main pulse (or TOAs) seen at the GMRT and the ASTROSAT in simultaneous observations. This is complicated by both time dependent and frequency dependent systematics in the TOAs. As is evident from \figurename{\ref{ortphsplt}}, the pulsar shows 
considerable timing noise, which is independent of frequency. As the  lower frequency TOAs are also affected by frequency dependent propagation effects, the timing noise was modeled using the regular cadence Fermi-LAT TOAs instead. These were fitted with a combination of eight sine waves in addition to the already fitted parameters in the reference ephemeris to model the red timing noise and obtain white  timing residuals using FITWAVES model in TEMPO2 \citep{hobbs2006tempo2}.

\begin{figure}[ht]
\includegraphics[scale=.23]{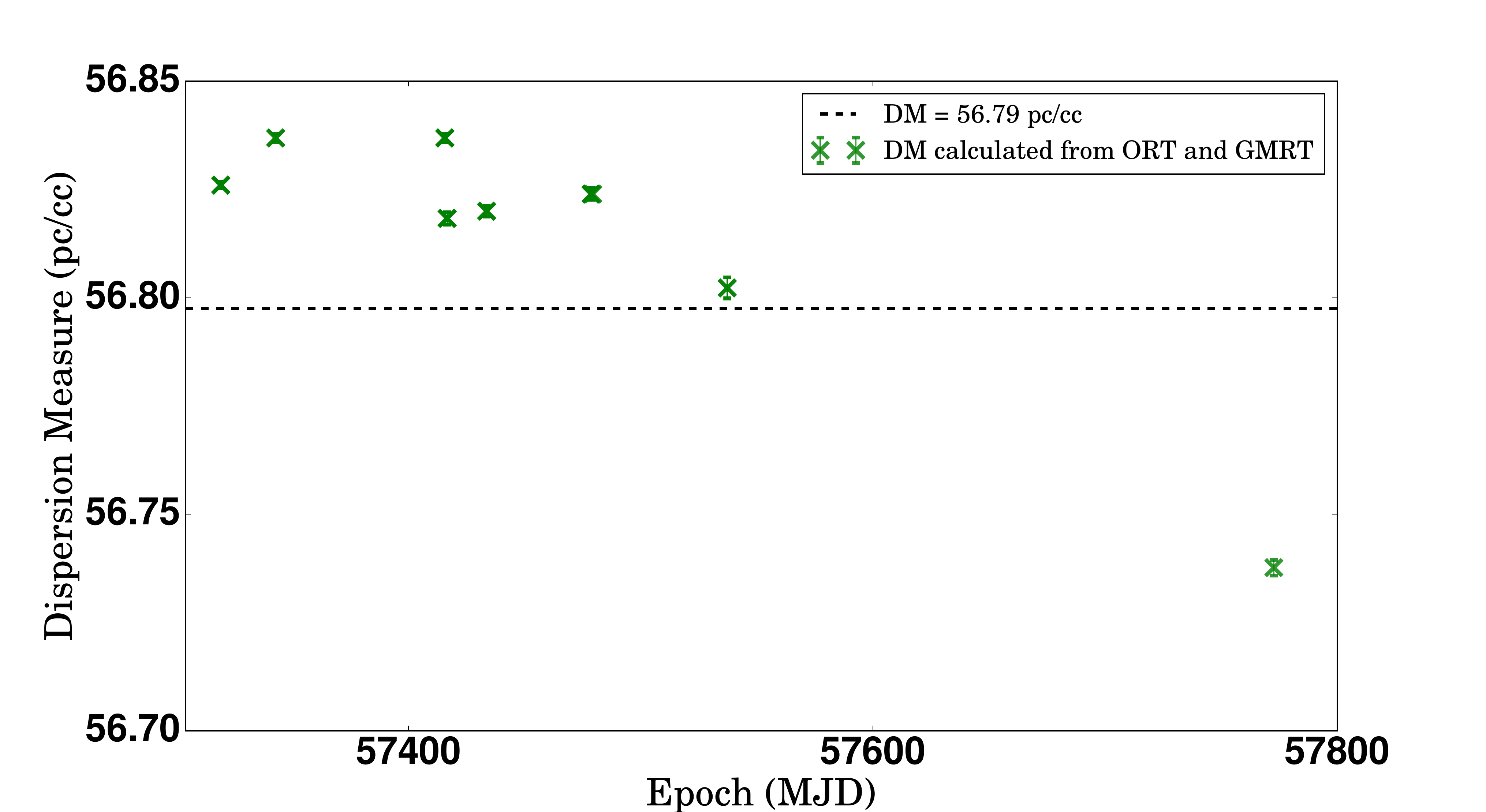}
\caption{Dispersion measure (DM) variations with observations epoch.}
\label{dmvar}
\end{figure}

As the pulsar is located in a dynamic pulsar wind nebula with 
nebular filaments, with trapped charged particles, moving 
across the line of sight, the DM of the pulsar and the pulse broadening
varies significantly from epoch to epoch. This 
introduces a systematic frequency dependent shift in barycentered 
TOAs, particularly significant for 
those derived from low radio frequency ORT topocentric TOAs. 
The typical variation in DM is on the order of 
0.01 pc\,cm$^{-3}$, which is equivalent to a shift of 21 $\mu$s and 
370  $\mu$s at 1390 MHz and 334.5 MHz respectively. Thus, it 
is essential to correct for DM variations to obtain 
reliable estimates for rotational parameters and lower 
post-fit timing residuals. We used the constrained DMMODEL 
in TEMPO2 to estimate the offsets from the chosen reference 
DM at epochs, where simultaneous ORT and GMRT observations were 
available. Our measurements are plotted in \figurename{ \ref{dmvar}}.

The DM model was used along-with the timing 
noise model and the astrometric and rotational model for PSR B0531+21 
for a fit to TOAs from the Fermi-LAT, the ASTROSAT-CZTI, 
the GMRT and the ORT. This corrects both the frequency independent 
and frequency dependent systematics in these TOAs allowing a 
more robust determination of relative offsets between the telescopes.

\section{Results and discussion}
\label{result}

\begin{figure}[h]
\includegraphics[scale=0.23]{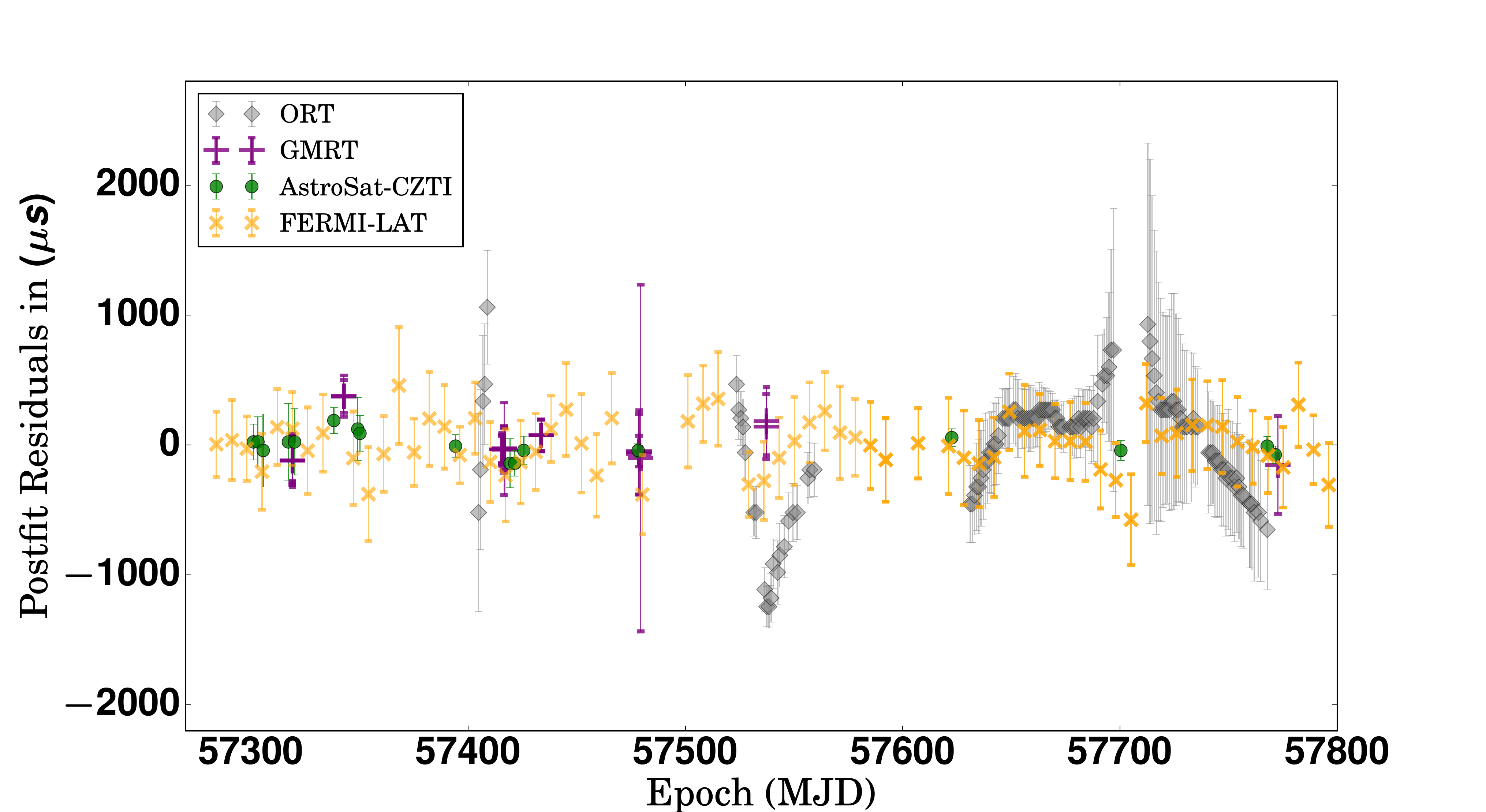}
\caption{Post-fit residuals after fitting the offsets between different telescopes. The symbols used are same as those used in \figurename{ \ref{ortphsplt}}.}
\label{postfitfinal}
\end{figure}

TEMPO2 provides a way to fit  the offsets between different telescopes 
and the resulting timing residuals are shown in Figure \ref{postfitfinal}.
\cite{rots2004absolute} concluded that the X-ray main pulse leads its radio 
counterpart by about 344 $\pm$ 40 $\mu$s. We can use this 
measurement to find out the ASTROSAT pipeline offset. The relative offsets 
between the GMRT and the CZTI aboard ASTROSAT was found to be 
-4716 $\pm$ 50 $\mu s$. While determining these offsets 
was the major objective of our project, we also determined in the process 
the offsets between the GMRT and the ORT and the GMRT and Fermi to be 
-29639 $\pm$ 50 $\mu s$  and -5368 $\pm$ 56 $\mu s$  respectively.
In addition, we verified that our timing solution fits the Jodrell Bank radio ToAs without introducing any time variable pattern.






The calibration of relative offsets between the radio and high energy 
emission is also important for a simultaneous radio $-$ high 
energy study of GPs to look for a radio$-$ high energy correlation. 
Such a study is currently underway.

\section{Acknowledgements}
This publication makes use of data from the Indian astronomy mission AstroSat, archived at the Indian Space Science Data Centre (ISSDC).
The CZT Imager instrument was built by a TIFR-led consortium of institutes across India, including VSSC, ISAC, IUCAA, SAC, and PRL. The Indian Space Research Organisation funded, managed and facilitated the project. We thank the staff of the Ooty Radio Telescope and the Giant Meterwave Radio Telescope for taking observations over such a large number of epochs. Both these telescopes are operated by National Centre for
Radio Astrophysics of Tata Institute of Fundamental Research. PONDER backend, used in this work, was built with TIFR XII plan grants 12P0714 and 12P0716.
We would like to thank the anominous referee for his/her useful comments and suggestions. AB 
would like to thank Alessandro Ridolfi for exposing to various techniques of PSRCHIVE package and Surajit Mondal for various fruitful discussions related to computational issues. We also thank Yogesh Maan for his valuable suggestions. BCJ, PKM and MAK acknowledges support for this work from DST-SERB grant EMR/2015/000515.

\bibliographystyle{aa}
\bibliography{offset_paper_ref}
\end{document}